\documentclass[11pt]{article}

\usepackage[centertags]{amsmath}
\usepackage{amsfonts}
\usepackage{amsthm}
\theoremstyle{remark}
\newtheorem*{remark}{Remark}
\usepackage{graphicx}
\usepackage{mathrsfs}
\usepackage{amssymb}
\usepackage{amsthm}
\usepackage{hyperref}

\begin{document}
\title{Modelling the Uruguayan debt \\through gaussians models}
\author{Andr\'es Sosa\footnote{Universidad de la Rep\'ublica, Facultad de Ciencias, Centro de Matem\'atica. Igu\'a 4225. 11400 Montevideo, Uruguay}\qquad Ernesto Mordecki\footnote{Universidad de la Rep\'ublica, Facultad de Ciencias, Centro de Matem\'atica. Igu\'a 4225. 11400 Montevideo, Uruguay}}

\maketitle
\begin{abstract}
We model bond's price curves corresponding to the sovereign u\-ru\-gua\-yan debt nominated in USD,  as an alternative to the official bond prices publication released by the Central Bank of Uruguay (CBU).  Four different gaussian models are fitted, 
based on historical data issued by the CBU,  corresponding to some of the more frequently traded bonds. The main difficulty we approach is the absence of liquidity in the bond market.  Nevertheless the adjustment is relatively good, giving the possibility of non-arbitrage pricing of the whole family of non traded instruments, and also the possibility of pricing derivative securities.
\end{abstract}
\section{Introduction}

Bond prices curves (or equivalently yield curves) constitute a major tool in debt analysis and perspective of the sovereign debt of a country.  Term-structure models have therefore been used in different ways by different classes of market participants. In the monetary policy context, the term structure is an indicator of the market's expectations regarding interest rate and inflation rates. From a financial point of view, the existence of a bond price curve helps the development of the domestic capital market, both for primary and secondary market.

There are essentially two approaches to model of the term structure. The general equilibrium approach starts from a description of the economy and derives the term structure of interest rate endogenously, as used for example by Cox Ingersoll and Ross in \cite{cir}. In contrast the arbitrage approach starts from assumptions about the stochastic evolution of one or more interest rates and derives prices of contingent claims by imposing the no arbitrage condition, this is, for example, the pioneering approach proposed by Vasicek \cite{v}.

In this first approximation to the problem of the Uruguayan USD nominated debt, we use the second approach and further we restrict our analysis to four gaussian models because of their analytical and numerical tractability. Factor models assume that the term structure of interest rate is driven by a set of ``state variables'' named as  ``factors''. As one factor generally explains a large proportion of the yield curve movement, it may tempting to reduce the analysis to one factor models. Nevertheless, the consideration of one factor implies the correlation between two different rates in the same time interval.  This perfect correlation is difficult to accept as empirical data usually shows high correlation,  so then we use  multiple factors model, and in particular, two factor models.

The first two models we choose are short rate models, the classical Vasicek model \cite{v}, and the more flexible G2++ model \cite{bm}. The parameter estimation is carried out based on maximum likelihood, following Chen and Scott \cite{cs}. The second couple of models belong to the HJM family \cite{hjm}.  Here we again choose first the simplest possible one,  the Ho-Lee model with constant volatility \cite{hl}, and second, the more flexible Hull-White model with tempered volatility \cite{hw}.  In this two last cases we calibrate the models with the help of minimization of squared differences.

Our contribution is in first place to provide an arbitrage-free set of bond prices for the Uruguayan USD debt, that can be used to portfolio valuation and derivative pricing.

The rest of the paper is organized as follows. Section 2 provides information about the Uruguayan debt and the publication of prices by CBU. Section 3 describes  the short rate model we use in the work and and the estimation methodology through maximum likelihood. In Section 4 we describe the results in the Vasicek and G2++ models. 
In Section 5  the HJM models are introduced, and the calibration procedure is explained.  Section 6 describes the results in Ho-Lee and Hull-White models, and Section 7 concludes.

\section{Bond price curves}

\subsection{About the Uruguayan debt}

The Uruguayan government issues debt through different financial instruments, 
with a variation of currencies and expirations.  In present times, the most relevant circulating instruments according to terms and currencies are
 
\begin{itemize}
\item the \emph{Treasury Notes}  in local currency and linked to CPI, 
\item the \emph{Local Bonds} (issued in the country)  linked to CPI (consumer price index) and in USD, 
\item the \emph{Global Bonds} (issued mainly in United States of America) in USD. 
\end{itemize}

\begin{figure}[h]
\centering \includegraphics[width=9cm,height=7cm]{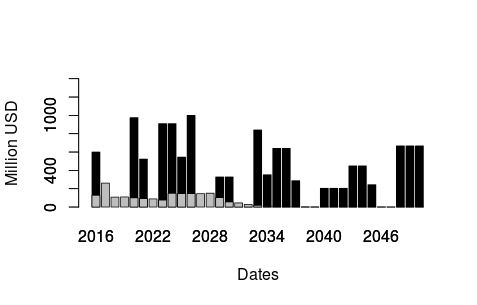}
\label{deudauru}
\caption{Central Government Debt Profile}
\end{figure}

To complete the analysis of the debt, it should be mentioned that the government also
holds obligations with multilateral financial institutions, estimated in less than the 10\% of the global debt amount. In Figure \ref{deudauru} we observe the  debt's profile of the Uruguayan debt across maturities.  The last expiration dates correspond to the 2050 bond, recently issued, paying its face value on thirds, in the years 2048, 2049 and 2050. 

Regarding the currency composition, more than one half of the Uruguayan debt is nominated in local currency (part of it linked to CPI), 45\% is issued in USD, and the rest corresponds to bonds issued in Euros and Yens. Almost all of the debt (94 \%) pays fixed coupons, the rest is mainly adjusted to Libor rate.

\subsection{Curve price estimation}

In most of the countries the corresponding regulating agencies and some financial corporations release yield curves corresponding to the sovereign debt.  The most popular methods include parametric methods (as the ones proposed by Nelson-Siegel \cite{ns} and Svensson \cite{sv}) also used non-parametric methods lesser extent.

Regarding Uruguay, the CBU does not release yield curves, but daily issues a report with prices for all the financial instruments issued by the government. The two authorized stock exchange institutions release their respective yield curves. The \emph{Electronic Stock Exchange of Uruguay (Bolsa Electr\'onica de Valores del Uruguay - BEVSA)} uses B-splines to produce  the yield curves corresponding to local currency, linked to CPI, and  to USD. The \emph{Montevideo Stock Exchange (Bolsa de Valores de Montevideo - BVM)} emits a single curve (linked to CPI) based on the methodology proposed by Svensson \cite{sv}.

The purpose of the present paper is to propose an alternative methodology for bond valuation,  for the debt issued in USD, to the one used by the CBU and also by the BVM and BEVSA. In the next subsection we briefly describe the procedure used by the monetary authority (CBU) to compute the prices of the different financial instruments.

\subsection{CBU pricing methodology}

The CBU issues daily reports containing prices of all the financial instruments issued by the Uruguayan government, named as ``price vector'' and available in the local stock exchanges. The pension funds and insurance companies, that are relevant participants in the domestic bond market, are obliged by law to use this prices in order to report portfolio values. 

This vector of prices is computed through a methodology that includes four different criteria, according to whether the bond has been traded or not, and according also to its expiration,  and apply with a hierarchical scheme. At the end of each business day the CBU releases information about bond prices, according to the following procedure:

\begin{enumerate}
\item For bonds that have been negotiated in the day (according to certain minimal amounts) the prices 
are computed as a weighted mean of the respective negotiation prices in the stocks exchanges. 
\item For bonds with less than a year to expire, an interpolation procedure between the previous price and the face value is applied to compute the prices
\item An \emph{index} $I(t)$ where $t$ represent the current  date is calculated in order to compute the prices of  the other bonds. This is a number that represent the value of an ``ideal'' mean bond. 
\item  For all other bonds, the new price is the previous price  multiplied by the ratio $\frac{I(t)}{I(t-1)}$ computed.
\end{enumerate}


The most relevant characteristic to be taken into account when analysing this procedure, is that the stock exchange of long expiration  debt instruments issued by the government (bonds) is really not liquid. This implies that most prices published by the CBU are computed instead of negotiated (see points 3 and 4 above), and this can happen for some instruments consecutively, during a relatively long period of time. This can lead to (theoretical) arbitrage opportunities, for instance giving larger yields for bonds with smaller maturities, what is equivalent to larger prices of zero coupon bond with smaller maturity than other zero coupon bonds. 
Of course, this arbitrage is eliminated in real negotiation, but is present, for instance, when evaluating portfolios. Although the intention of the methodology is to follow the movements of the market,  one should always take into account this fact.

\section{Short rate models}

When modelling yield curves through stochastic processes, the classical approach consists in modelling the short rate through a certain amount of sources of uncertainty, under the denomination of ``states''. When this states are used to construct the short interest rate we obtain a large variety of models, described in a vast part of the literature in fixed income mathematical finance \cite{bjork},  \cite{bm}, \cite{fili} and \cite{musiela}.

In this part of the work we use two gaussian models, the one proposed by Vasicek in 1977 \cite{v}, that is seminal in the literature. The second model we apply is the G2++ \cite{bm}, that is a modification the previous one that has two factors and includes the initial price curve, avoiding in this way arbitrage at the initial time. This model has in fact a more general version, as it can include an arbitrary amount of factors, and is called the Gn++ model \cite{gn} in this case. However, the choice of the number of factors then involves a compromise between numerically efficient implementation and capability of the model to represent realistic correlation  patterns and to fit satisfactorily enough market data.  

In the next two subsection we review each of the models to be used.

\subsection{Vasicek's model}

In his  classical paper \cite{v}, Vasicek proposes a  model for the short rate through a stochastic differential equation driven by a Wiener process,
\begin{equation*}
dr(t)=a(b-r(t))dt+\sigma dW(t)\qquad\qquad r(0)=r_0;
\end{equation*} 
where $a$, $b$ y $\sigma$ are positive constants and $W(t)$ is a standard Wiener process.  The solution to this equation is know as the Ornstein-Uhlenbeck process.  
It defines an elastic random walk around a trend, with a mean reverting characteristic. Given the set of information at time $s$, the short rate $r(t)$ is normally distributed with
\begin{align*}
\mathbf E\: (r_t|\mathcal{F}_s) =& r_se^{-a(t-s)}+b\Big(1-e^{-a(t-s)}\Big)\\
Var\: (r_t|\mathcal{F}_s) =& \frac{\sigma^2}{2a}\Big(1-e^{-2a(t-s)}\Big).
\end{align*}

The bond price can be obtained by computing the discounted expected terminal value of the bond with respect to a risk neutral probability measure $Q$. This quantity can be explicitly computed concluding that the Vasicek model  is an affine model  whose the solution is
\begin{equation}
\label{vasicekprecio}
P(t,T) = A(t,T)e^{-B(t,T)r_t};
\end{equation}
where 
\begin{align*}
A(t,T)&= \exp \Big( \big(b-\frac{\sigma^2}{2a^2}\big) (B(t,T)-T+t) - \frac{\sigma^2}{4a}B(t,T)^2\Big);\\
B(t,T)&= \frac{1}{a} \Big(1-e^{-a(T-t)}\Big).
\end{align*}

\subsection{The G2++ model}

As mentioned above, the price correlation given by the Vasicek model to different instruments is 1. For this reason, as observed prices do not show such high correlation, emerged the idea to introduce models with more factors, that imply more parameters, that would allow to better fit to the data. One of this proposals is the Gn++ model, proposed in \cite{gn} that we use with two factors. Another relevant characteristic of this proposal is that in takes into account the whole initial price curve.

The dynamics of the short rate  process in this model is given by
\begin{equation*}
r(t)=x(t)+y(t)+\varphi(t), \hspace{1cm}r(0)=r_0;
\end{equation*}
where the process $x(t)$ y $y(t)$ and the vector $(W_1,W_2)$ satisfy
 \begin{align*}
 &dx=-ax dt+\sigma dW_1;\\
 &dy=-by dt+\eta dW_2;\\
 &dW_1(t)dW_2(t)=\rho dt;
 \end{align*}
where $r_0, a, b,\sigma, \eta$ are positive constant and $-1\leq \rho \leq 1$. Given the set of information at time $s$, the short rate $r(t)$ is normally distributed with
\begin{align*}
E(r(t)|\mathcal{F}_s)=& x(s)e^{-a(t-s)}+y(s)e^{-b(t-s)}+\varphi(t);\notag\\
Var(r(t)|\mathcal{F}_s)=&\frac{\sigma^2}{2a}(1-e^{-2a(t-s)})\notag\\
+&\frac{\eta^2}{2b}(1-e^{-2b(t-s)})+\frac{2\rho\sigma\eta}{a+b}(1-\exp^{-(a+b)(t-s)}).
\end{align*}

The price at time $t$ of a zero coupon bond with maturity at time $T$ is
 \begin{multline}
 \label{g2precio}
 P(t,T)=\exp \Big(-\int_t^T \varphi(u)du-\frac{1-e^{-a(T-t)}}{a}x(t)\\
 -\frac{1-e^{-b(T-t)}}{b}y(t)+\frac{1}{2}V(t,T)\Big);
 \end{multline}
where
 \begin{align}
 \label{V}
   V(t,T)=&\ \frac{\sigma^2}{a^2}\Big(T-t+\frac{2}{a}e^{-a(T-t)}-\frac{1}{2a}e^{-2a(T-t)}-\frac{3}{2a}\Big)\notag\\
  &+\frac{\eta^2}{b^2}\Big(T-t+\frac{2}{b}e^{-b(T-t)}-\frac{1}{2b}e^{-2b(T-t)}-\frac{3}{2b}\Big)\notag\\
  &+\frac{2\rho\sigma\eta}{ab}\Big(T-t+\frac{e^{-a(T-t)}-1}{a}+\frac{e^{-b(T-t)}-1}{b}-\frac{e^{-(a+b)(T-t)}-1}{a+b}\Big).
 \end{align}

Even though the previous formula \eqref{g2precio} gives bond prices in the model, it is necessary to estimate the function $\varphi$. In order to do this, it is necessary to assume that the initial price curve $T\mapsto P^M(0,T)$ is known. The model fits the currently observed term structure if $P^{Model}(0,T)=P^M(0,T)$, therefore the price at time $t$ of a zero coupon bond maturity at time $T$ is
\begin{equation}
\label{PBCg2}
P(t,T)=\frac{P^M(0,T)}{P^M(0,t)}\exp(A(t,T));
\end{equation}
where
\begin{equation*}
A(t,T)=\frac{1}{2}\big( V(t,T)-V(0,T)+V(0,t)\Big) -\frac{1-e^{-a(T-t)}}{a}x(t)-\frac{1-e^{-b(T-t)}}{b}y(t).
\end{equation*}

\subsection{Maximum Likelihood Estimation}

Observe that in formulas \eqref{vasicekprecio} and \eqref{PBCg2}, the log-prices are expressed as a linear function of the state variables. In order to determine this state variables in both models it is necessary to  use one bond price time series in Vasicek model and two time series in G2++.  But to carry our this procedure the values of the parameters (to be estimated) is necessary. This is why we use the Maximum Likelihood (ML) method with the help of a change of variables, that gives the corresponding Jacobian term in the ML expression. In both models, based on the Markov property of the processes, the joint density is written as a product of conditional densities, each of one has normal distribution, with parameters three parameters in the first case ($a,b,\sigma$), and five in the second ($a,b,\sigma,\eta,\rho$).

Given the panel data set $P_t^M=\big(P^M(t,T^{(i)})\big)$, $i=1,...,I$ y $t=1,...,T$, where $P^M(t,T^{(i)})$  is the price at time $t$ of the zero coupon bond with maturity $T^{(i)}$ and denote by  $X_t=\big(x_t^{(i)}\big)$, $t=1,...,T$ the state vector.

The joint density of $P_2^M,...,P_T^M$ satisfies  
\begin{equation*}
f(P_2^M,...,P_T^M|P_1^M,\Theta)=\prod_{k=2}^{T}f(P_k^M|P_{k-1}^M,\Theta)
\end{equation*}
by the Markov property. Changing variables in each conditional density, we obtain
\begin{align*}
f(P_k^M|P_{k-1}^M,\Theta)	&=f(X_k|X_{k-1},\Theta) |\frac{\partial X_k}{\partial P_k}|\\
								&= f(X_k|X_{k-1},\Theta) |\frac{1}{\frac{\partial P_k}{\partial X_k}}|= f(X_k|X_{k-1},\Theta)\frac{1}{|J_k|};
\end{align*} 
where $J_k$ is the Jacobian of the change of variables.

 \begin{remark}
We find adequate in our situation to use as many bond price time series as number of factors. But this is not strictly necessary. In some cases, when more time series than state variables are used it is possible to introduce additional random variables in order to estimate the model \cite{cs}.
\end{remark}

\section{Short rate estimation}

\subsection{Data}

The information used in the construction of the yield curves is the one provided by the CBU, for USD nominated bond prices, traded both in the domestic market an in foreign exchanges. It should be noticed that, besides the liquidity problem of the bond market, there are no derivatives on this instruments, so all the available information is provided by the bond prices.

The data to be used in the estimation procedure corresponds to the time period  January 4, 2010 to October   30, 2013. In the Vasicek model we use the bond \textit{BE330115P}, expiring in January 2033, and in the G2++ model we use the same one an add the bond \textit{BE250928F} with expiry in September 2025. We choose these bonds as they are the more frequently traded. Each time series is processed according the coupon payment scheme (amount and frequency) to obtain the corresponding yields, that give the zero coupon bond prices used in the estimation.

\subsection{Estimation in the Vasicek model}

\begin{figure}[h]
\centering \includegraphics[width=9cm,height=5cm]{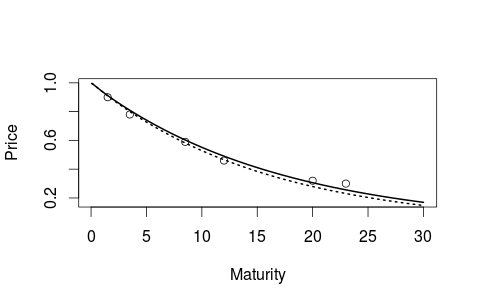}
\label{vasicek}
\caption{The solid line shows the first estimation with Vasicek model,  the dotted one the second (that uses only negotiated prices). The dots correspond to market prices of zero coupon bonds. } 
\end{figure}

In order to evaluate the influence of the non liquid computed prices (as explained above) we proceed in this case estimating the parameters corresponding to the Vasicek model in two situations. We first estimate parameters using all the available information, i.e. we use weekly bond prices taken on Wednesdays. The obtained parameters are:
$$
a=1.7051; \qquad b=0.0937; \qquad \sigma=0,3721.
$$
In the second case we use only negotiated prices (504 observations). This implies that the time intervals between prices are not regular, leading to a slightly more complicated estimation scheme. The obtained parameters are
$$
a=1,7145;\qquad b=0,0896;\qquad \sigma=0,4971.
$$
This comparison help us to verify that the variation in the values of the parameters is not significant (perhaps the most important variation is registered in $\sigma$) and for this reason in what follows, and in this first approach to the problem, we will use all the available time series. In Figure \ref{vasicek} we observe the price curve corresponding to August 13, 2013 given by the model in both estimations carried out, and also the bond prices issued by the CBU.

\subsection{Estimation in the G2++ model}

As we mentioned above we use two bond prices time series, with expiration in 2025 and 2033 respectively.  We take weekly prices corresponding to Wednesdays. The obtained values with the ML estimation described above are
$$ 
a=0.1300;\quad b=0.3526;\quad \sigma=0.2062;\quad \eta=0.4892; \quad\rho=-0.99.
$$ 
Using this values with the corresponding bond prices we can obtain the daily time series corresponding to the short rate. This allow us to compute the bond prices for arbitrary maturities in each one of the days. We show the bond price curve for the following maturities: 1, 2, 3, 6 and 9 months and 1, 2, 3, 5, 7, 10, 15, 20 and 25
years.

\begin{figure}[h]
\centering \includegraphics[width=9cm,height=8cm]{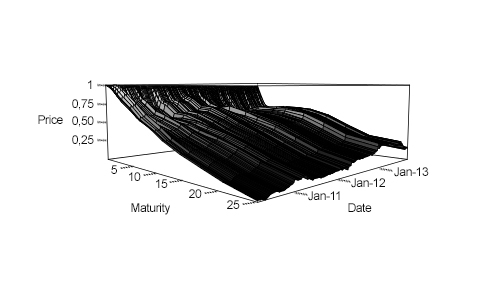}
\label{3}
\caption{Surface of zero coupon bond prices using the G2++ model}
\end{figure}

In Figure \ref{3} we see the daily zero coupon bond price curve corresponding to the analysed time period issued by the Uruguayan government, as a result of the application of the G2++ model.

If we analyse the data in Figure \ref{3} for the second semester of 2012, the model gives a curve with non-negative slope for some maturities. This fact introduces arbitrage possibilities, as it gives cheaper bonds with  smaller maturities than others with larger maturities.  More precisely, our model adjustment give parameter that result in increasing bond prices in some intervals of the maturity  $T$. Our parameter set includes a correlation very close to $-1$, and instances of negative values of the factors $x(t)$ and  $y(t)$.  We will return to this fact in more detail in the next subsection.

\subsection{On arbitrage possibilities in the G2++ model}

We analyse how formula \eqref{PBCg2} for a zero coupon bond price in G2++ model can give arbitrage opportunities.
Equation  \eqref{V} can be written as
\begin{multline*}
V(t,T)=\frac{\sigma^2}{a^2}\int_t^T(1-e^{-a(T-u)})^2du+\frac{\eta^2}{b^2}\int_t^T(1-e^{-b(T-u)})^2du\\
+\frac{2\rho\sigma\eta}{ab}\int_t^T(1-e^{-a(T-u)})(1-e^{-b(T-u)})du.
\end{multline*}
We then take logarithms and differentiate formula \eqref{PBCg2} w.r.t $T$, to obtian
\begin{align*}
\frac{\partial}{\partial T} P(t,T)&=\frac{\partial}{\partial T} P^M(0,T)+\frac{\sigma^2}{2a^2}\Big((1-e^{-a(T-t)})^2-(1-e^{-aT})^2\Big)\\
&\quad+\frac{\eta^2}{2b^2}\Big((1-e^{-b(T-t)})^2-(1-e^{-bT})^2\Big)\\
&\quad+\frac{\rho \sigma\eta}{ab}\Big((1-e^{-a(T-t)})(1-e^{-b(T-t)})-(1-e^{-aT})(1-e^{-bT})\Big)\\
&\quad- e^{-a(T-t)}x(t)-e^{-b(T-t)}y(t).
\end{align*}
In this formula the first three addends are negative. 
The fourth has the opposite sign of $\rho$, the last two depend on $x(t)$ and $y(t)$.
In our application  $\rho$ is close to $-1$, what is associated with the fact that $x(t)$ and $y(t)$ take opposite signs, therefore, in some times intervals the derivative is positive, as shown in Figure \ref{3}. Nevertheless the model gives general valuable information about the structure of the debt.


\section{Heath-Jarrow-Morton model}

In the market we do not have a real instantaneous interest rate. In certain cases, 
the one (or three) month interest rate series is used as a proxy to estimate this interest rate.  It is not convenient to use the overnight rate, as it has very high volatility due to economical factors, as for instance liquidity, and this can not depend on the structure of the yield curve.

By this reason, and also to avoid arbitrage opportunities in a systematic way, Heath, Jarrow and Morton introduce a new methodology \cite{hjm} (refereed as HJM models), that models the \emph{forward instantaneous rate} at time $t$ by a stochastic differential equation driven by a Wiener process
\begin{equation*}
f(s,t)=f(0,t)+\int_0^t\alpha(u,t)du+\int_0^t\sigma(u,t)dW_u;
\end{equation*} 
where $W_t$ is a Wiener process.

We depart from the free arbitrage model, that assumes that  the 
$\alpha(u,t)$, $0\leq u\leq t$ and $\sigma(u,t)$, $0\leq u\leq t$  are adapted processes defined in an underlying stochastic basis $(\Omega,\mathcal{F},\mathcal{F}_u,Q)$, and $Q$ is a risk-neutral probability measure. The key aspect of HJM techniques lies in the recognition that the drifts of the no-arbitrage evolution of certain variables can be expressed as functions of their volatilities and the correlations among themselves. 
\begin{equation}\label{eq:drift}
\alpha(s,t)=\sigma(s,t)\int_s^t\sigma(s,u)du.
\end{equation}

Therefore, in order to specify an HJM model the initial forward rate curve $f(0,t)$ and the volatility structure $\sigma(s,t)$  should be given because  no drift estimation is needed. It should be observed that as long as the function $\sigma$ is deterministic,  by condition \eqref{eq:drift} the drift is also deterministic, in consequence, the forward rates are gaussian.

In the applications that follow, we first choose the simplest possible alternative for the volatility, i.e. we assume $\sigma(s,u)=\sigma>0$  is a positive constant. This gives
\begin{equation*}
f(s,t)=f(0,t)+\sigma(st-s^2/2)+\sigma W_t;
\end{equation*}
the so called Ho-Lee model \cite{hl}. The price of zero coupon bond in this model is then given by
\begin{equation}\label{phl}
P(t,T)=\frac{P^M(0,T)}{P^M(0,t)}\exp\Big((T-t)f^M(0,t)-\frac{\sigma^2}{2}t(T-t)^2-(T-t)r(t)\Big).
\end{equation}

In the second model we assume that $\sigma(s,u)=\sigma e^{-a(u-s)}$, where $a$ and $\sigma$ are positive constants.  This gives
\begin{equation*}
r(t)=f(0,t)+\frac{\sigma^2}{2a^2}(e^{-a t}-1)^2+\sigma\int_0^t e^{-a(t-u)}dW_u.
\end{equation*}  

The price of a zero coupon bond in this case is
\begin{multline}
\label{phw}
P(t,T)=\frac{P^M(0,T)}{P^M(0,t)}\exp\Big(B(t,T)f^M(0,t)\\
-\frac{\sigma^2}{4a}(1-e^{-2t})B(t,T)^2-B(t,T)r(t)\Big).
\end{multline}
that corresponds to the Hull-White model \cite{hw} with time dependent parameters.

\section{Results for HJM models}

In order to use it in practice to price contingent claims, we have to calibrate its parameters on market data. To do this we can adopt  a cross-sectional approach. Assume we observe a cross-section of market prices of contingent claims, that is to say, the prices of a set of $N$ zero coupon bonds made by the market at the same day. For time $t$, let us denote the vector prices  $P_t^M=\big(P^M(t,T^{(i)})\big)$, $i=1,...,I$  where $P^M(t,T^{(i)})$ is the price in time $t$ the zero coupon bond maturity at time $T^{(i)}$ and assume we are able to compute the price vector  made by the model and denote them by $P_t^{\small{Model}}=\big(P^{\small{Model}}(t,T^{(i)})\big)$, $i=1,...,I$. Model prices are functions of the parameter vector  of the model, see \eqref{phl} in Ho-Lee and \eqref{phw} in Hull-White.

The idea is to find the values for  $\Theta$ minimizing the difference between market
prices and model prices. To do this we have to solve the  least square problem,
\begin{equation*}
\min_{\Theta}\frac{1}{I}\sum_{i=1}^{i=I}\Big(P^M(t,T^{(i)})-P^{\small{Model}}(t,T^{(i)})\Big)^2.
\end{equation*}

In this analysis we use weekly prices corresponding to the whole year 2014, 
of 10 USD nominated bonds issued by the CBU. The maximum expiration date corresponds to the \textit{BE451120F} bond, in November 2045. To obtain an approximation of the initial interest rate we used the reference curve \textit{CUD-BEVSA} daily issued by BEVSA for three months, and to adjust the initial forward rate curve we used the yield curve corresponding to January 5, 2013.

\begin{figure}[h!]
\centering \includegraphics[height=4cm]{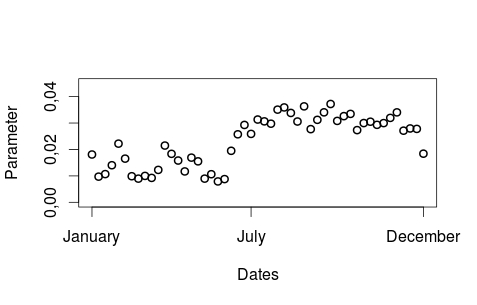}
\caption{Calibrated Ho-Lee $\sigma$ parameter over time.}
\label{hla}
\end{figure}

Calibrating the Ho-Lee model we obtain a $\sigma$ for each day.  In Figure \ref{hla} we observe then the daily variation of $\sigma$.  The mean value of the estimation is $0.0232$, with a standard deviation of $0.0094$.  

\begin{figure}[h!]
\centering \includegraphics[scale=0.40,height=4cm]{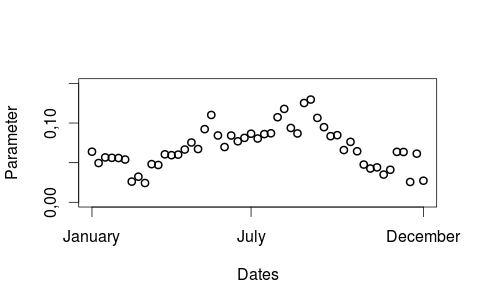}
\centering \includegraphics[scale=0.40,height=4cm]{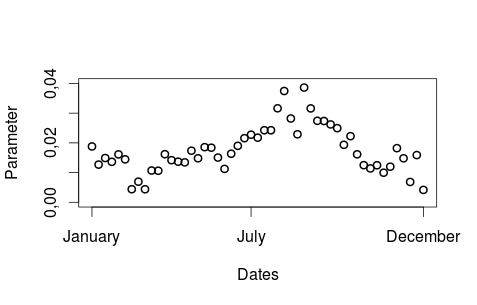}
\caption{Calibrated Hull-White parameters over time. On the left $a$, on the right $\sigma$.}
\label{hw}
\end{figure}

For the Hull-White model the same procedure is carried out.  In the Figure \ref{hw} (left) we observe the daily values of $a$ and in Figure \ref{hw} (right) the values of $\sigma$. The mean value for $a$  is $ 0.0693$ with a standard deviation of $0.0257$. 
The corresponding mean and standard deviation for $\sigma$ are $0.0177$ and $0.0079$ respectively.

\begin{figure}[h]
\centering \includegraphics[scale=0.40,height=3.8cm]{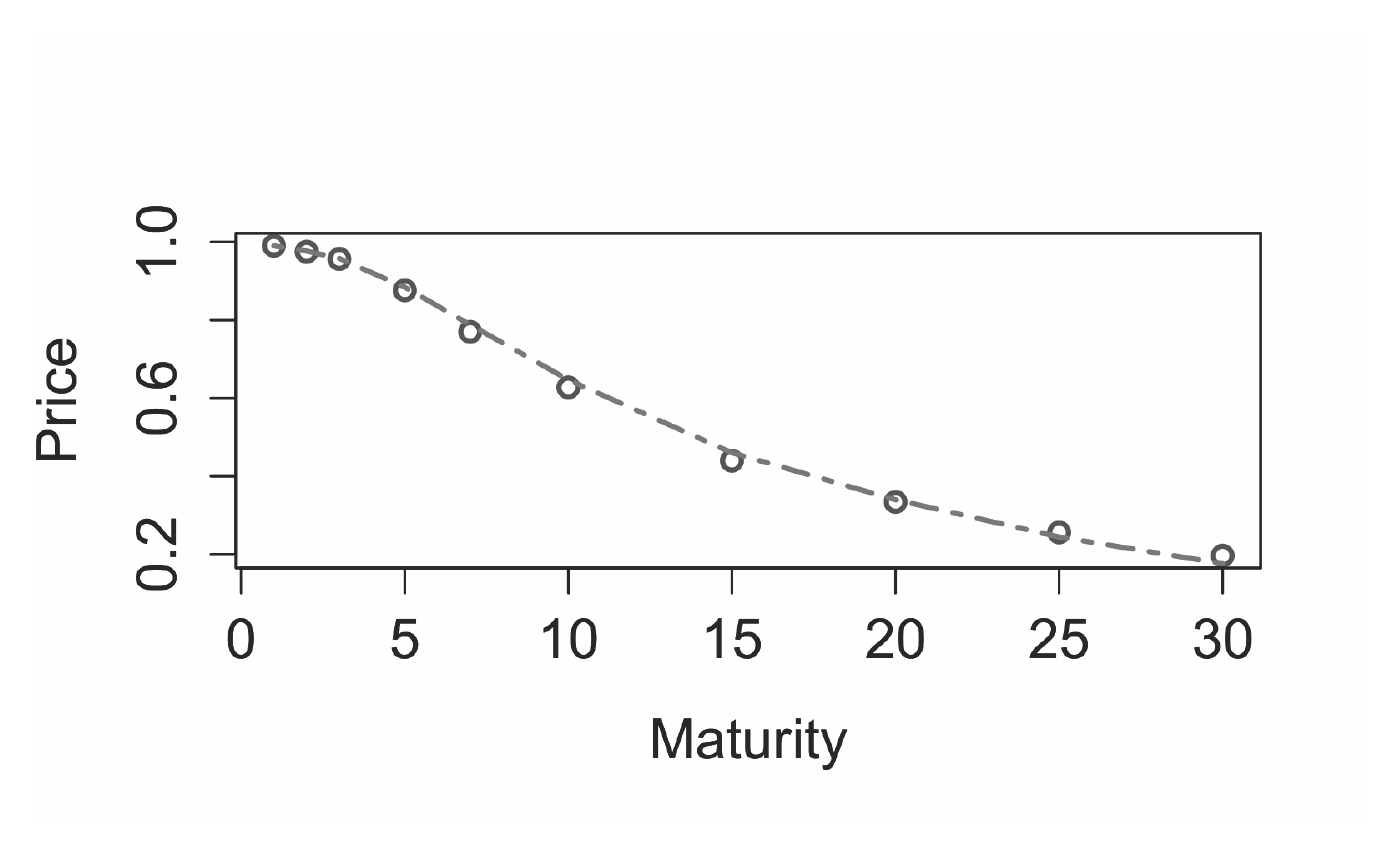}
\centering \includegraphics[scale=0.40,height=3.8cm]{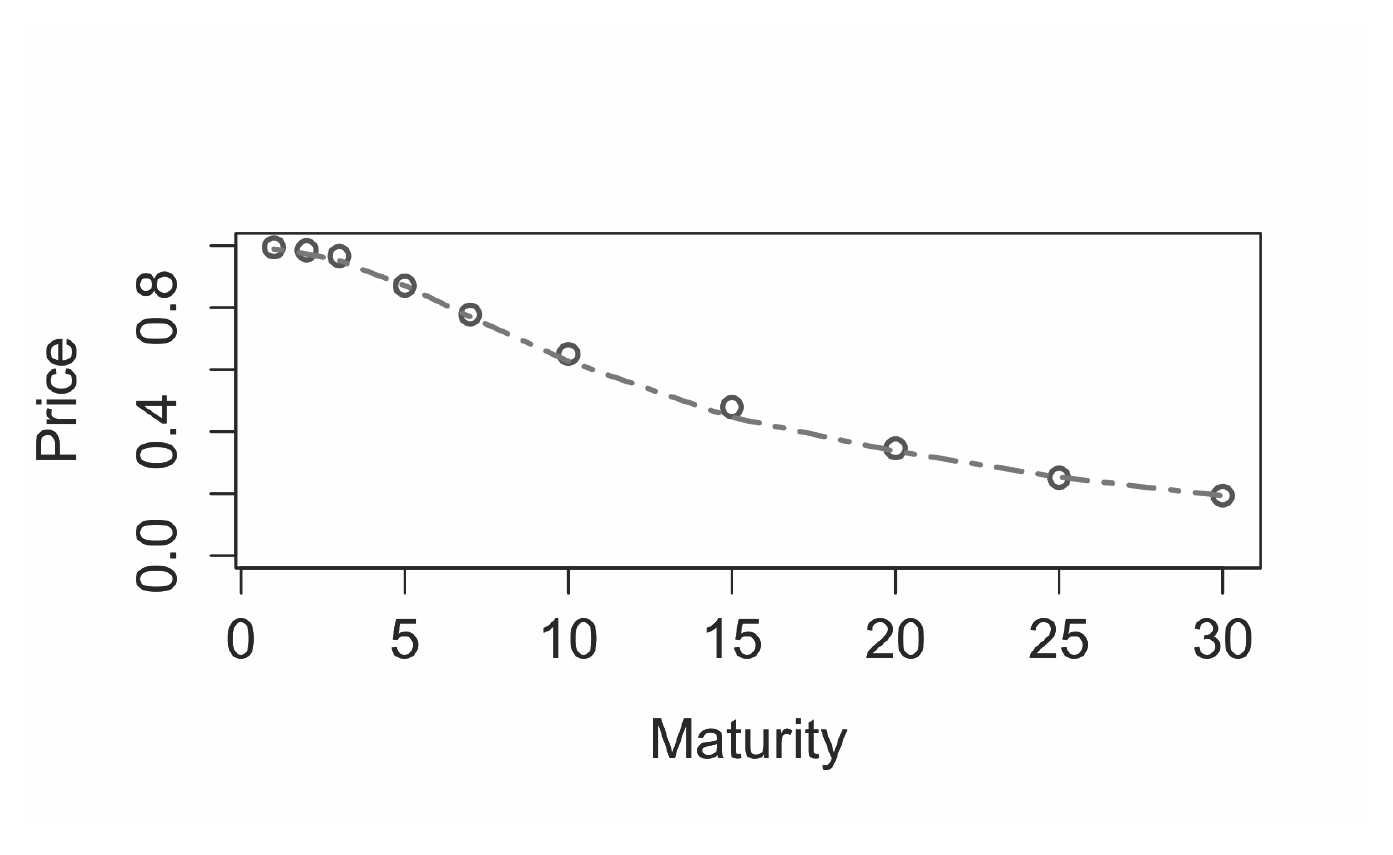}
\caption{Bond price curve adjusted to market data. On the left Ho-Lee, on the right Hull-White. In both cases the dots represent the market prices.}
\label{grhw}
\end{figure}

The calibration of the Ho-Lee model corresponding to Wednesday 25th of June of 2014 gives $\sigma=0.3071$,  and is shown in  Figure \ref{grhw} (left). In  Figure \ref{grhw} (right) we see the calibration for the same day corresponding to the Hull-White model. The parameters in this case are $a=0.0813$ and $\sigma=0.0215$. 

\begin{figure}[h]
\centering \includegraphics[width=12cm,height=8cm]{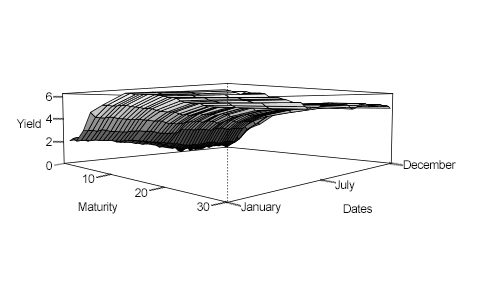}
\caption{Term structure in Uruguayan Market via Hull-White Model.}
\label{ts}
\end{figure}

Both in the Ho-Lee and  Hull-White models, the daily calibration allow us to construct term structure of interest rates for the corresponding day. The result is shown in Figure \ref{ts}.

\section{Conclusions}

The objective of the present work is to present an arbitrage-free consistent model to price the  Uruguayan debt nominated in USD. A second purpose is to provide an instrument capable of pricing derivatives on bonds, as interest rate swaps that are beginning to be used in the Uruguayan market. Based on data from the Uruguayan market for coupon bonds we adjust four different gaussian models.

This information is of valuable interest to financial practitioners and policy makers alike. Policy makers  monitor expectations of future monetary policy to gauge the effectiveness of their  strategy. For practitioners, the availability of accurate interest rate forecasts can be the key to a successful trading strategy. We hope that this modelling exercises  would enrich our understanding of market expectations and improve the characteristic behaviour of the term structure of interest rate. 

First we adjust the Vasicek and the G2++ models for interest rates, departing from data from one bond and two bonds respectively. We follow the methodology proposed by Chen and Scott \cite{cs} that departs from historical data to estimate the parameters of the model through maximum likelihood. The second model is more flexible than the first one, as it involves more parameters, and also adjust the initial interest rate curve avoiding the possibility of arbitrage. Nevertheless, as we show, when high values of correlation are used (and this is our case), the model can give arbitrage situations. 

Second, we model the debt through the HJM models. We choose two different situations, assuming first that the volatility is constant, giving raise to the Ho-Lee model, the simplest HJM model, and second we assume a exponentially decreasing volatility, giving rise to the two parameter Hull-White model. In this case the adjustment seems to be better, as no arbitrage possibilities appear (the HJM model uses the initial curve and also works under a risk-neutral measure), and the different values for different days in both cases show certain stability. We conclude that HJM models fit better to the data, and are capable of describing the whole market structure with information of a series of then more traded bonds. 

Further work includes the consideration of more sophisticated HJM models, some test to verify the adjustment of issued bond prices that are not used in the calibration procedures, and the computation of derivatives prices in bonds.

\end{document}